\documentclass{iau} 
\usepackage{graphicx}

\title[S 341.~~The SEDs of AGNs] 
{The Spectral Energy Distributions of Active Galactic Nuclei}

\author[Brown et al.]   
{M. J. I. Brown,$^{1}$
K. J. Duncan,$^{2}$
H. Landt,$^{3}$
M. Kirk$,^{1}$
C. Ricci$^{4,5,6}$ and 
N. Kamraj$^{7}$}

\affiliation{$^1$School of Physics \& Astronomy, Monash University,\\ Clayton, Victoria 3800, Australia \\ email: {Michael.Brown@Monash.edu}
\\[\affilskip]
$^{2}$Leiden Observatory, Leiden University,\\ NL-2300 RA Leiden, Netherlands
\\[\affilskip]
$^{3}$Department of Physics, Centre for Extragalactic Astronomy, Durham University,\\ South Road, Durham DH1 3LE, UK
\\[\affilskip]
$^{4}$N\'ucleo de Astronom\'ia de la Facultad de Ingenier\'ia, Universidad Diego Portales,\\ Av. Ej\'ercito Libertador 441, Santiago, Chile
\\[\affilskip]
$^{5}$Kavli Institute for Astronomy and Astrophysics, Peking University,\\ Beijing 100871, China
\\[\affilskip]
$^{6}$Chinese Academy of Sciences South America Center for Astronomy,\\ Camino El Observatorio 1515, Las Condes, Santiago, Chile
\\[\affilskip]
$^{7}$Cahill Center for Astronomy and Astrophysics, California Institute of Technology,\\ Pasadena, CA 91125, USA
}

\pubyear{2015}
\volume{xxx}  
\jname{Title of your IAU Symposium}
\editors{A.C. Editor, B.D. Editor \& C.E. Editor, eds.}
\begin{document}

\maketitle

\begin{abstract}
We present ongoing work on the spectral energy distributions (SEDs) of active galactic nuclei (AGNs), derived from X-ray, ultraviolet, optical, infrared and radio photometry and spectroscopy. Our work is motivated by new wide-field imaging surveys that will identify vast numbers of AGNs, and by the need to benchmark AGN SED fitting codes. We have constructed 41 SEDs of individual AGNs and 80 additional SEDs that mimic Seyfert spectra. All of our SEDs span 0.09 to $30~{\rm \mu m}$, while some extend into the X-ray and/or radio. We have tested the utility of the SEDs by using them to generate AGN photometric redshifts, and they outperform SEDs from the prior literature, including reduced redshift errors and  flux density residuals.
\end{abstract}

\firstsection 
\section{Introduction}

Photometric redshifts and AGN SED modelling will be critical for understanding the vast numbers of AGNs identified by wide-field surveys with new facilities such as eROSITA, ASKAP and LOFAR. It is also increasingly common to include AGN components in galaxy SED modelling, so empirical AGN SEDs could also prove critical for benchmarking the performance of SED modelling codes. 

\begin{figure}
\begin{center}
 \includegraphics[width=0.80\textwidth]{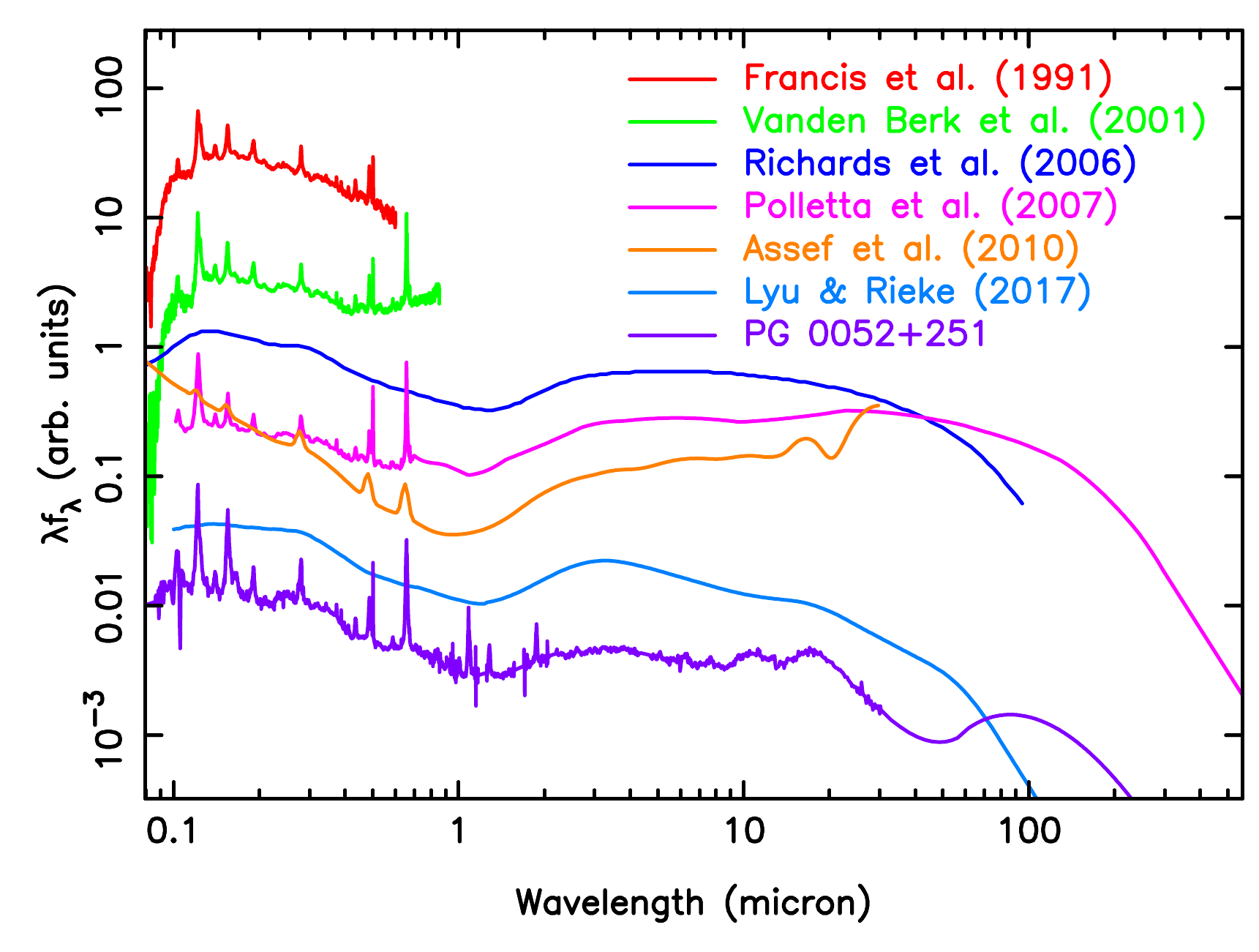} 
\end{center}
 \caption{Quasar SEDs from the past three decades \cite[(Francis et~al. 1991, Vanden Berk et~al. 2001, Richards et~al. 2006, Polletta et~al. 2007, Assef et~al. 2010, Lyu \& Rieke 2017)]{francis1991,vandenberk2001,richards2006,polletta2007,assef2010,lyu2017} along with our new SED for PG~0052+251. The trend towards greater spectral resolution and wavelength coverage is evident. Our PG~0052+251 SED has broad wavelength coverage while including near-infrared emission lines and 2mid-infrared silicate emission features. 
   \label{fig:iauqso}}
\end{figure}

AGN SEDs have been developed by various groups over the past three decades, and some illustrative examples of quasar SEDs from the literature are provided in Figure~\ref{fig:iauqso}. The SEDs in Figure~\ref{fig:iauqso} were developed with a variety of data, goals and methods, so some caution is required when comparing them. That said, there is an overall trend towards improved spectrophotometric accuracy, wavelength coverage and spectral resolution.

For this work we have produced AGN SEDs by combining X-ray, ultraviolet, optical, infrared and radio spectroscopy and photometry of individual objects (Brown et al., submitted). This approach can exploit the expanded wavelength range and spectrophotometric accuracy of ground-based telescopes and satellites from the past decade. However, it comes with the risk that variability and wavelength-dependent (extraction) aperture bias will produce unrealistic SEDs. To mitigate this risk we compare our SEDs with photometry and utilise the SEDs to determine AGN photometric redshifts.

\section{Constructing AGN SEDs}

To produce SEDs of individual AGNS we combine ground-based optical and near-infrared spectra with space-based X-ray, ultraviolet and infrared spectra. The availability (or lack) of near-infrared spectrophotometry strongly limits the building of SEDs, particularly as the near-infrared can include significant contributions from the AGN disk, torus and host galaxy. We mostly use published spectra or reduced spectra provided by archives, with only a few exceptions (NuSTAR, VLT XShooter, VLT SINFONI). 

To rescale and verify spectrophotometry, and constrain SED models, we have measured aperture photometry across the ultraviolet to mid-infrared wavelength range using images from GALEX, Swift, SDSS, PanSTARRS, Skymapper, 2MASS and WISE. At longer wavelengths we have used published photometry from {\it Herschel}, WMAP, {\it Planck} and single-dish radio telescopes. 

Most of the AGNs in our sample do not have far-infrared spectra, but the shape of far-infrared spectrum is sufficiently simple that it can often be adequately modelled with a grey-body curve. In the radio, we have modelled SEDs using power-laws and polynomials, which is adequate to model the observed flux densities with an accuracy of $\sim 0.1~{\rm dex}$. Polynomials are used to interpolate over gaps in spectral coverage, including {\it Akari} to {\it Spitzer} and gaps in near-infrared spectra caused by atmospheric absorption. 

\begin{figure}
\begin{center}
 \includegraphics[width=0.78\textwidth]{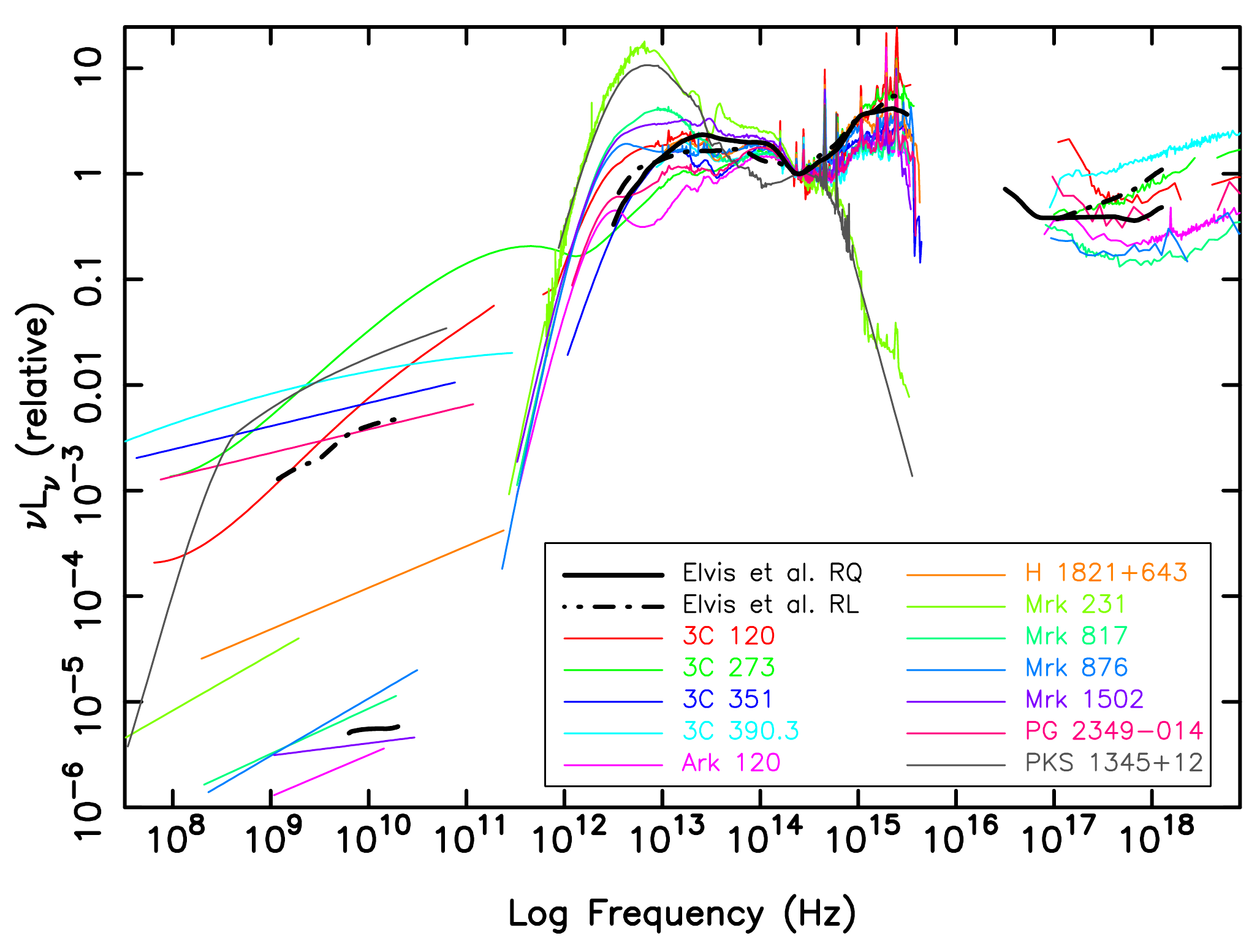} 
\end{center}
 \caption{The AGN SEDs from our library with radio coverage, normalised at $1.25~{\rm \mu m}$, along with the \cite[Elvis et~al. (1994)]{elvis1994} radio-loud and radio-quiet quasar SEDs for comparison. The diversity of AGN SEDs is apparent, including the broad range of radio luminosities and the diversity of SED shapes in the mid and far-infrared.}
   \label{fig:elvis}
\end{figure}

To produce continuous SEDs, the individual spectra are initially scaled using the matched aperture photometry. The multiplicative scaling is then adjusted so neighbouring spectra have consistent flux densities. AGN variability and wavelength dependent (extraction) aperture bias does result in some SEDs being unphysical and inconsistent with the photometry, and these are rejected from our sample. For most of the remaining 41 AGNs, the multiplicative scalings range between 0.3 and 3 in the ultraviolet, and converge towards 1 with increasing wavelength. All of the SEDs span the 0.09 to $30~{\rm \mu m}$, but a subset extend into the X-ray, far-infrared and/or radio. Examples of several of our SEDs, along with the \cite[Elvis et~al. (1994)]{elvis1994} radio-loud and radio-quiet quasar SEDs, are shown in Figure~\ref{fig:elvis}.

Our 41 SEDs for individual AGNs are biased towards luminous quasars and the central regions of nearby Seyferts, where the light is dominated by the AGN and aperture bias is thus reduced. To approximate the SEDs of Seyferts, including host galaxy light, we mix the SEDs of the central regions of Seyferts with galaxy SEDs taken from the \cite[Brown et al. (2014)]{brown2014} sample, resulting in 80 additional SEDs spanning 0.09 to to $30~{\rm \mu m}$.

\section{Photometric redshifts}

We have tested the utility of our AGN SEDs by using them (and other SED libraries) to generate photometric redshifts (photo-$z$s) for X-ray selected AGNs in the Bo\"otes. We compare photo-$z$s generated with our SEDs against those generated using the \cite[Brown et~al. (2014)]{brown2014} SEDs, which don't include powerful AGNs, and the AGN SEDs of \cite[Ananna et~al. (2017)]{ananna2017}, which build upon the \cite[Polletta et~al. (2007)]{polletta2007} SEDs and were recently used to produce photo-$z$s for X-ray selected AGNs in Stripe-82. The photo-$z$s were determined using the EAZY code \cite[(Brammer et al. 2008)]{brammer2008} run on optical, near-infrared and mid-infrared photometry using the methods of \cite[Duncan et al. (2018)]{duncan2018}.

\begin{figure}
\begin{center}
 \includegraphics[width=0.95\textwidth]{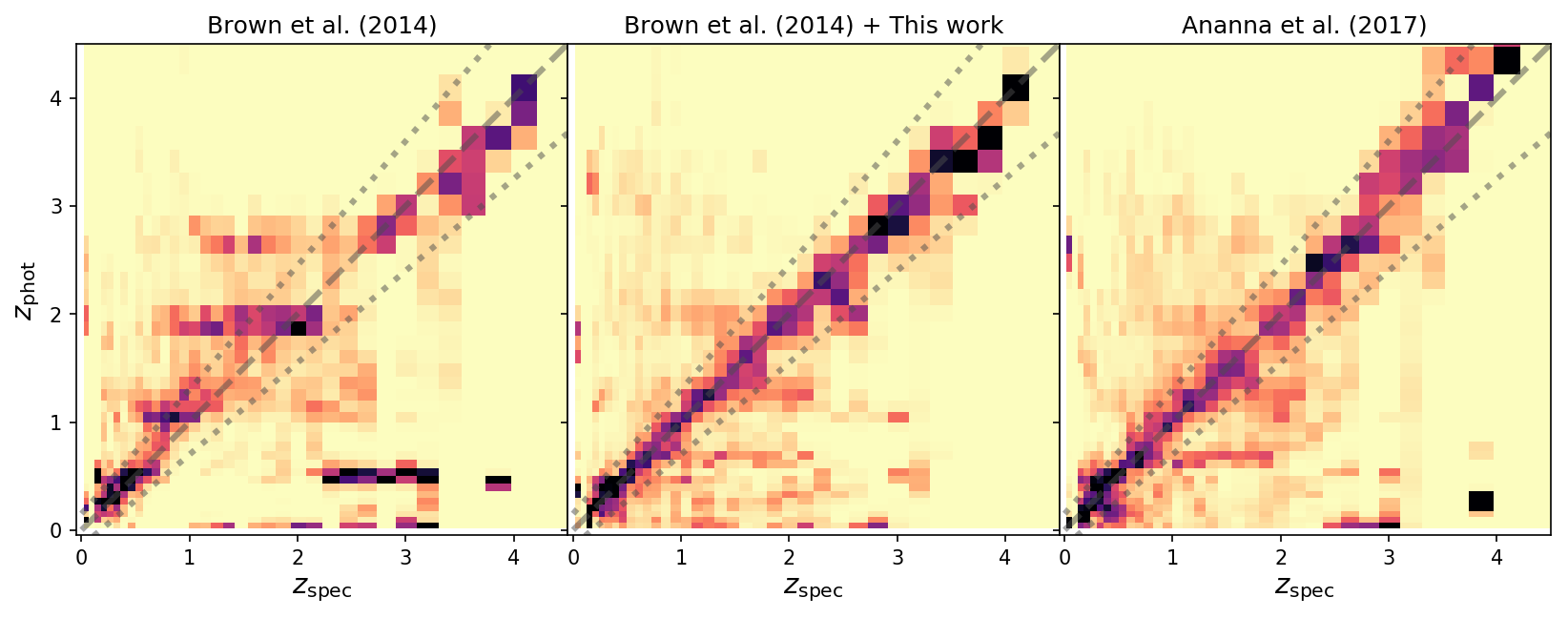} 
 \caption{X-ray AGN photometric redshifts as a function of spectroscopic redshift, determined with EAZY while using different SED libraries. The plots are colour coded using the sum of the redshift probability density functions, so dark regions correspond to photometric redshifts with small uncertainties. Our new SEDs, in combination with the \cite[Brown et~al. (2014)]{brown2014} galaxy SEDs, produce the highest quality photometric redshifts, particularly for $z>1.5$ quasars.}
   \label{fig:photoz}
\end{center}
\end{figure}

Figure~\ref{fig:photoz} illustrates the performance of the various SED libraries. Unsurprisingly, the \cite[Brown et~al. (2014)]{brown2014} galaxy SED library, which lacks powerful AGNs, performs poorly with $\sigma_{\textup{NMAD}}\sim 0.8 \times (1+z)$ for $z\sim 3$ quasars. Our new SEDs, in combination with the \cite[Brown et al. (2014)]{brown2014} galaxy SEDs, produce significantly better quality photo-$z$s than the previous library alone and shows improvements over photo-$z$s using the \cite[Ananna et~al. (2017)]{ananna2017} SEDs. Our photometric redshifts have a typical  scatter of $\sigma_{\textup{NMAD}}=0.096 \times (1+z)$ and  flux density residuals are $\simeq 9\%$, with the exact value varying with wavelength. We attribute this improvement in part to the availability of new spectra, including from {\it Akari} and {\it Spitzer}. 
A complete and representative library of AGN SED templates will be vital for producing template-based photo-$z$s that are comparable in precision/reliability to those produced through the latest empirical methods \cite[(Duncan et al. 2019)]{duncan2019}. Such redshifts will be necessary for as new AGN surveys push into parameter space that currently lacks the spectroscopic training samples required for empirical photo-$z$ estimates.  

\begin{discussion}

\discuss{David Rosario}{Is there much variation in UV-optical SED shape?}

\discuss{Michael Brown}{For the broadline quasars there's not that much variation in the continuum shape, although strong emission lines do contribute to the overall SED. There's significantly more variation in the mid-infrared, including the presence and absence of dust components.}

\discuss{David Rosario}{What is the approach to the SED hybrids?}

\discuss{Michael Brown}{We take the central SED of a Seyfert and then add contributions from a host galaxy SED. For example, 25\% Seyfert and 75\% host galaxy (defined at a wavelength of $0.6~{\rm \mu m}$).}

\discuss{Themiya Nanayakkara}{What are the differences in performance between the empirical SEDs and models?}

\discuss{Michael Brown}{When the new SEDs are fitted to AGN photometry they typically produce a reduced $\chi^2\sim 1$, which is promising. This is, in part, due to the inclusion of strong emission lines, which are often difficult to include in SED models.}

\discuss{Themiya Nanayakkara}{Can you explain the $z=0$ photometric redshift outliers?}

\discuss{Michael Brown}{All of the SED libraries produce some $z=0$ photometric redshift outliers, and in some instances there are highly variable objects with photometry that wildly varies with wavelength, resulting in high $\chi^2$.}

\end{discussion}

\end{document}